# Consideration of Learning Orientations as an Application of Achievement Goals in Evaluating Life Science Majors in Introductory Physics

Andrew J. Mason[1] and Charles A. Bertram[1,2]

[1]*University of Central Arkansas, Department of Physics and Astronomy, Conway, AR 72035*

[2]*Cabot High School and Cabot Freshman Academy, Cabot, AR 72023*

**Abstract:** When considering performing an IPLS course transformation for one's own institution, life science majors' achievement goals are a necessary consideration to ensure the pedagogical transformation will be effective. However, achievement goals are rarely an explicit consideration in physics education research topics such as metacognition. We investigate a sample population of 218 students in a first-semester introductory algebra-based physics course, drawn from 14 laboratory sections within six semesters of course sections, to determine the influence of achievement goals on life science majors' attitudes towards physics. Learning orientations that respectively pertain to mastery goals and performance goals, in addition to a learning orientation that does not report a performance goal, were recorded from students in the specific context of learning a problem-solving framework during an in-class exercise. Students' learning orientations, defined within the context of students' self-reported statements in the specific context of a problem-solving-related research-based course implementation, are compared to pre-post results on physics problem solving items in a well-established attitudinal survey instrument, in order to establish the categories' validity. In addition, mastery-related and performance-related orientations appear to extend to overall pre-post attitudinal shifts, but not to force and motion concepts or to overall course grade, within the scope of an introductory physics course. There also appears to be differentiation regarding overall course performance within health science majors, but not within biology majors, in terms of learning orientations; however, health science majors generally appear to fare less well on all measurements in the study than do biology majors, regardless of learning orientations.

## I. INTRODUCTION: ACHIEVEMENT GOALS IN PER

### A. Achievement goal theory and potential applications to introductory physics

One perhaps overlooked aspect of cognitive psychology within the physics education research community, especially in terms of considering attitudes towards learning physics, is the consideration of student achievement goals,[1] specifically in terms of being a motivating factor for learning class material.[2,3] Achievement goals have been defined in cognitive literature as self-reported reasons for how and why people engage in achievement situations.[4,5] A long-featured means of treating achievement goals regards a contrast between two sets of goals: mastery goals, e.g. genuine interest in mastering an aspect or skill within the lesson's content (e.g. problem solving skills), and performance goals, e.g. trying to achieve the best grade or get the most points possible within the course.[6] Achievement goals include well-researched cognitive topics, e.g. approach-avoidance motivations, which have a long history of prominent focus within the realm of psychology[3] and have been studied in the context of both mastery goals and performance goals.[7] Concerns about the effect of a classroom environment's achievement goals on developing student motivations to succeed, throughout the K12 and college/university-level curriculum, have also arisen in recent years.[7-9] Mastery and performance goals may also affect other aspects of student learning outside the classroom; e.g. students who are motivated by mastery of the material have been found to be more likely to successfully transfer knowledge.[5]

### B. Achievement Goal Theory: Lack of Explicit Consideration in PER

While efforts to address student motivation in PER deserve recognition,[10] the explicit study of achievement goals' effect on student motivation, as discussed above, appears to remain largely unexplored within the PER community. Furthermore, motivation is recognized as being a complex topic to analyze, and efforts to properly consider it in terms of cognition and meta-cognition tasks have been relatively recent.[10,11] This is an important topic to address, however, as recent evidence suggests that student motivation may affect research of other aspects of physics education, e.g. problem solving[10] or teacher motivation and teacher content knowledge.[12]

A rare example of explicit consideration of achievement goals is the treatment of "learning orientations" by Hazari et al.,[13] pertaining to a longitudinal study of graduate students in the physical sciences as they finished their terminal degrees and began their careers. Hazari et al. found that students who were more "learner-oriented" (i.e. they were primarily motivated to learn physical science content and skills out of genuine interest and appreciation for the material's inherent value) were more successful on average than were students who were more "performance-oriented" (i.e. more focused on performing well and attaining a well-defined benchmark for success). The "learner" and "performance" orientations in the Hazari et al. study strongly reflect mastery vs. performance achievement goals as discussed within cognitive and educational psychology communities.[3,4]

## II. BACKGROUND: IPLS CONSIDERATIONS AND LEARNING ORIENTATIONS

### A. IPLS: An Implicit Consideration of Achievement Goal Theory

Implicitly, recent trends in PER show promise in identifying and addressing student motivations. A recent example in physics education research entails course transformations in university-level Introductory Physics for Life Sciences (IPLS) courses. IPLS courses address a recognized need to form a stronger interdisciplinary curriculum that integrates STEM disciplines.[14-17] An overarching consideration for IPLS course designers is the need to transfer physics content knowledge in terms of pedagogical needs of life science majors.[18-28] This issue has long been understood in application to introductory physics courses with predominantly life science majors.[29-31] IPLS courses reported in recently literature are typically designed for specifically biology majors,[24-26] or for another specific monolithic group of life science majors, e.g. pre-physical therapy majors.[28] In IPLS literature, a frequent aim is to focus on introductory physics topics that can have a direct application to pertinent life science topics for a specific course track for the life sciences. Authors of IPLS courses typically must consult with life science faculty members to determine what topics are more important to emphasize within a biological context. In addition, one must address certain skills acquired while learning physics (e.g. problem solving[25] or laboratory skills[26]) in the context of life science applications.

The focus on the application of physics topics in the context of a biology major track may be regarded as an implicit treatment of life science majors' achievement goals. Educational goals of biology majors, pre-physical therapy majors, and other life science majors are necessarily a consideration in recently reported IPLS course transformations, requiring instructors to explicitly describe and analyze their pedagogical goals.[20]

In a study from Redish and Cooke,[23] biologists and physicists revealed differences in perspectives on introductory physics, to the point of expressing cultural and epistemological differences. For example, introductory biology classes do not emphasize problem solving and quantitative reasoning skills in the same way that introductory physics classes do. If the faculty of the two respective departments have such deeply different perspectives on introductory physics, a similar epistemic gap exists between physical science majors and life science majors, unless introductory physics instructors take steps to demonstrate the relevance of the course material to life science.[32]

Epistemic views are thus an implicit consideration of achievement goals. Demonstration of the relevance of physics concepts in modeling biological systems, and the relevance of skills such as problem solving to biological applications,[33] are a means of emphasizing the importance of mastering the material, as opposed to merely performing well in the course and/or satisfying requisite courses.

### B. Learning Orientations within Different Sets of Life Science Majors

The pertinence of achievement goals within IPLS course reforms is important to consider when analyzing biology majors as well as health science majors in an introductory physics class prior to beginning course reforms. In previous studies,[34-36] the authors of this paper studied attitudinal shifts, gains in conceptual understanding on force and motion, and in-class group problem solving habits for an introductory algebra-based physics course sections. The research-based implementation that students had to work on entailed a weekly in-group problem solving exercise. The exercise entailed a live attempt to solve a context-rich problem,[37-39] followed by an individual metacognitive exercise in diagnosing one's strengths and weaknesses in solving the problem, with the use of a rubric adapted from a study on self-diagnosis of quiz mistakes.[40-41] The course population is typically dominated by life science majors, and so the researchers' long-term pedagogical aims have been to gather background information for the purposes of IPLS-style course development, as described in the previous section, with particular attention paid to problem solving and teamwork skills.

However, the life science majors within the course population were split into two approximately equally sized groups: biology majors, who reside within the same natural science college as did the department of physics and astronomy that offered the course; and health science majors, who reside within a separate health science college. Because the life science majors are effectively divided into two groups along these lines (namely, natural science biology majors as opposed to health science majors), the question is raised whether there are different learning experiences in the same introductory physics course between these two groups of life science majors.[42] For example, the physics course may have to address different content needs for both sets of life science majors. Another example is that the two sets of majors may have, on average, different achievement goals in mind, regarding learning problem solving skills within an introductory physics context.

Achievement goals became a necessary consideration in analyzing feedback survey data from an initial study during the Spring 2014 semester.[34] The survey question that prompted further analysis in the initial studies was initially intended to investigate which steps of a problem-solving framework students felt they were most helped in developing by the problem-solving exercise. (Please see Fig. 3 for a verbatim copy of the survey's introduction and two principal questions.) Approximately a third of student respondents (deemed "Framework-oriented") answered the question in terms of their perceived inherent value of all or part of a problem-solving framework. (e.g. "it helped me learn how to gather information from the problem" or "it helped me understand how to approach the problem step-by-step"). Another third of student respondents (deemed "Performance-oriented"), on the other hand, answered the question in terms of performance (e.g. how the exercise helped them do their homework, study for their exams, and/or prepare them for the lab activity that followed the exercise). Here, the two orientations of Hazari et al. ("learning-oriented" and "performance-oriented") become apparent, albeit within a much smaller scope of a single course. More generally, the result pointed to the treatment of mastery vs. performance achievement goals within cognitive psychology literature.

However, the remaining third of student respondents (deemed "Vague-oriented") did not answer in terms of achievement goals. This group of students either discussed the problem-solving process itself (i.e. that they liked working in groups and getting feedback from partners) or answered in a way that was not pertinent to the survey question (e.g. complaints about the homework, or a brief assertion that the exercise was useful without explanation).

Initial analysis of learning orientation categories[34] appeared to show a relationship between learning orientation and choice of major. Biology majors from the Spring 2014 section appeared to be much more likely to be framework-oriented than did health science majors. In addition, biology majors tended to perform better in the overall course, experience higher gains on FCI[43] and experience more expert-like attitudinal shifts in CLASS[44] data, than did health science majors. While the initial analysis appeared to show a correlation between learning orientation and choice of major in terms of quantitative performance, sample size within a single course section was too small to make any definitive conclusions about the general course population. Similarly, analysis of in-class audiovisual data,[35] while permitting the researchers to identify variables, did not offer enough of a sample size to make any determinations of relationship between learning orientation and in-class group performance and dynamics.

## C. Research Questions of the Current Study

Recognizing that the self-reported data from student feedback surveys in previous studies lies within given definitions for achievement goals,[4,5] the authors have continued the study with data collected from six semester sections of the introductory algebra-based physics course, ranging from the Spring 2014 to Spring 2017 semesters, for a total of over 200 students. This group includes approximately equal numbers of health science and biology majors. The sample also includes a smaller population of other natural science majors and a small number of non-science majors; analysis, however, will focus on life science majors, using data from non-life science majors primarily as an approximate reference point.

The research questions of this paper are within the context of exploring subsets of the featured student sample as described above. Questions #1 and #2 directly relate to the validity of the learning orientations, and questions #2 and #3 relate to factors outside the learning orientations that may prove to be more pertinent than, and/or unrelated to, learning orientations.

1) Students are grouped in terms of self-expressed achievement goals in the context of a problem solving exercise within an introductory physics course, e.g. a similar categorization of learning orientations to Hazari et al.[13] Can a form of validity be established via comparison to pre-post normalized gains within an already-validated attitudinal instrument (namely, the CLASS survey[44])?
2) If the premise in question 1 is at least provisionally established, to what extent do different learning orientations map to variance in measurements across the entire course?
3) To what extent does the difference between a natural science curriculum and a health science curriculum affect course outcomes, and is this difference related with learning orientations?

The justifications for these research questions are as follows.

### 1. Validity of the Learning Orientation Categories between Students: Pre-Post Attitudinal Shifts on Problem Solving

First, as in previous literature by the researchers, a serious concern arises regarding the validity of the aforementioned learning orientation categories. While cognitive psychology literature appears to indicate that such a classification is robust and rich in applications to other aspects of cognition, we specifically consider using this classification within the smaller scope of an introductory physics course, in which learning orientations may not make much difference. Therefore, the first research question entails a means of establishing validity for the learning orientations used in the study.

Our chosen method is to establish a form of convergent construct validity using a comparison to an already-validated instrument. Specifically, the learning categories, as defined by students' self-expressed statements, should reflect pre-post attitudinal shifts on a well-established survey instrument such as the CLASS. Ref. 6 (p. 122) concisely summarizes the difference between mastery goals and performance goals: "Students with mastery orientation seek to improve their competence. Those with performance orientations seek to prove their competence." This suggests an expectation that Framework-oriented students, who have mastery-related achievement goals, will be more likely to have positive expert-like gains on attitudes towards physics problem solving than will the Performance-oriented (i.e. expressing performance goals) students. As students' learning orientations in this study are established upon survey response data that pertained specifically to a research-based problem-solving exercise, analysis of pre-post attitudinal shifts on the problem-solving item clusters of the CLASS should demonstrate a more expert-like shift for Framework-oriented students than for Performance-oriented students. If on the other hand there is no difference between the two learning orientations on CLASS problem solving item shifts, the result would instead suggest that the learning orientation categories are not sufficiently valid for the scope of a single introductory physics course.

### 2. Comparison between Learning Orientations on General Measurements for the Entire Course

Second, we examine whether specific results between learning orientations on the CLASS problem solving item clusters may extend to more general pre-post CLASS results (both overall and on other specific item clusters), as well as whether similar differences appear in pre-post FCI results as well as in overall course performance. It does not necessarily follow that learning orientations, as defined within the framing of a specific aspect of the course, will show similar relative patterns in terms of the entire course or other aspects of the course. However, students who express a mastery motivation to learn have been shown to more readily transfer learning to other topics.[5] This finding suggests that if a specific orientation shows more expert-like shifts on attitudes towards problem solving in physics, then it may extend to more expert-like shifts on overall attitudes towards physics, and perhaps for that matter performance-related results. We therefore examine the learning orientations on more general pre-post CLASS results, pre-post FCI results, and course performance.

### 3. Degree of Influence of Learning Orientations in Comparisons between Biology and Health Science Majors

Third, we consider pre-post FCI and CLASS comparisons, as well as overall course grade comparisons, between biology majors and health science majors, and address whether learning orientations influence this comparison. Specifically, we consider whether any of the three different learning orientations show differences between biology majors and health science majors. If it turns out that biology majors and health science majors' relative differences cannot be explained by learning orientations, then the difference between both types of life science majors must be considered as a distinct separate variable regarding course performance and pre-post diagnostic analysis. If on the other hand the respective learning orientations within the group of biology majors exhibit different behaviors than do their counterparts within health science majors, then it will suggest that the two sets of majors will have a qualitative difference in terms of achievement goals, which will influence any relative performance differences between two sets of majors.

## III. METHODOLOGY

### A. Student Population and Course Structure

The student population for the host institution's first-semester introductory algebra-based introductory physics course is predominantly a mixture of biology majors and health science majors (typically 80-90% for any given section); the average semester's section enrollment contains an approximately equal number of each kind of life science major. The university in question was a primarily undergraduate state university; however, the biology department also houses a master's program, and the health science college houses a master's program, a Ph.D. program, and a professional school for physical and occupational therapy. Biology majors and health science majors who wished to enroll in graduate or professional programs, whether at the host institution or elsewhere, could be expected to express a form of achievement goal, whether to master the physics material or to perform well in the course.

Data was collected from fourteen total laboratory sections of this course, covering six semesters ranging from Spring 2014 to Spring 2017. Students who failed the overall course, were absent for class on any day for which data was taken, or noticeably failed to take at least one of the pretest and posttest surveys seriously, were omitted from the pool of data at the end of each semester. The resulting student sample contained 218 total students. 91 students in this sample were biology majors and 85 were health science majors. The remainder of the student population consisted mostly of other Natural Science majors, in the same college as the Biology majors (35 students), who also needed the course for their respective majors' requirements. Most of the other Natural Science majors were either computer science majors or chemistry majors. Seven non-science majors also took the course to satisfy a general education requirement for physical science.

> You are studying a herd of bighorn sheep in Alberta, and have found two rams butting heads with each other to establish dominance. Your camera is able to get a high-speed recording of the rams as they take a running start toward each other, on level ground and along a straight line path, before they collide head-on. This way you can use video analysis software to determine how fast each ram was going before the collision. The first ram looks to be about average size for an adult, while the second ram looks to be about ½ the size of the first ram, so you estimate the mass of the first ram to be about 90 kg* and the second ram to be about 45 kg. In your analysis, you see that the second ram was running much faster (2.10 m/s) than the first ram was (0.90 m/s), such that when they butted heads, the smaller ram was brought to a halt. As a result, you find that you can determine the velocity of the larger ram after the collision.

**FIG. 1**. Example of a context-rich problem provided during the laboratory group problem solving exercise.

**B. Research-Based Problem-Solving Exercise**

The course was in a traditional lecture-laboratory format with no recitation section. Therefore, in order to explore a recitation-oriented innovation, the first 60 minutes of the laboratory section was reserved for a research-based pre-laboratory exercise, consisting of a preliminary cooperative group problem solving exercise[37,39] on context-rich problems,[38] to be followed by the scheduled experiment for the remainder of the laboratory period. In this way, the researchers could focus upon students' learning orientations for a specific metacognitive activity within the course.

Each laboratory group consisted of 2-3 students in each of eight laboratory groups, with a maximum of 24 students per laboratory section. The exception to this situation was in the Spring 2017 semester, which used a newly built laboratory room that contained only 6 laboratory tables with 4 students each. To compensate for this, the Spring 2017 students were instructed to first work in pairs on the problem, then consult the other pairing at the same table. Each week's problem focused upon a central conceptual element that also featured in the regular laboratory exercise that followed the problem-solving exercise. The background of each problem was framed in a real-world situation, related to either biology or health science as could be accommodated by the concepts that pertained to the following laboratory activity. Fig. 1 shows an example of a context-rich problem given for this exercise.

| Circle how much you think you understood on: | What were your strengths on each of these parts? | What did you struggle with on each part? |
|---|---|---|
| **Problem description**<br><br>Full/Partial/None | | |
| **Solution construction**<br><br>Full/Partial/None | | |
| **Logical progression**<br><br>Full/Partial/None | | |

**FIG. 2**. Rubric used by students for metacognition, in consideration of a generalized problem-solving framework.

The students took approximately 30-50 minutes to solve the problem (depending upon how quickly a particular problem could be finished). The groups could ask for assistance from the instructor as needed, and were permitted use of textbooks and notebooks. (The latter measure was necessary because a significant number of students arriving underprepared for the exercise or otherwise being unable to proceed, to ensure that students in each group could work with the same knowledge base.) Conceptual follow-up questions were also provided so that groups who finished the main problem quickly could continue working as other groups finished the main problem.

| | |
|---|---|
| This survey is to gather information about certain activities within the course this semester. The reflection exercise at the beginning of the lab is intended to assist in problem solving techniques on homework assignments. The exercise focused on such skills as: <br> - Gathering information from the problem (knowns and what wasn't known) <br> - Choosing a way to solve the problem from the notes <br> - Setting up a solution process <br> - Checking over your work <br> 1) In what ways did you find this exercise useful towards learning the material in the course? <br> 2) Do you have any suggestions to make this exercise more useful toward learning the material in the course? | |

| Orientation | Sample Responses to Question #1 |
|---|---|
| Framework (citing a problem solving step on which student was helped; discussing problem solving framework as a whole; relating problem solving to other aspects of course) | I thought the exercises were very useful in ***increasing my understanding of certain topics***. They also helped me learn ***how to approach different situations***. [Spring '15, biology major] |
| | It was useful b/c it taught you to always ***list your knowns and unknowns***, and how important it is to ***draw your problem*** so you can ***visualize*** what's going on. [Fall '16, health science major] |
| | It was useful to see the ***concepts we learned in class applied to problems***. This is how I learned and comprehended material. [Spring '17, health science major] |
| Performance (mentioning how exercise helped study for exams, work on homework or the laboratory exercise) | It helped give an exercise, using the information ***we were going to cover in lab that day*** [Spring '15, health science major] |
| | Very useful, gives insight on ***how class material could be presented during tests***, and are more helpful than worked examples from class just because we spend more time. [Fall '16, biology major; answer to Question #2: ***"Incentives? Like bonus."***] |
| | It helped us think through scenarios on our own, which ended up ***preparing us for the test better***. [Spring '17, biology major] |
| Vague (comments on perceived benefits of working in groups; vague description of benefits of exercise; non-detailed assertion that the exercise was useful) | It was helpful ***to see how other people approached the problems***. [Spring '15, biology major] |
| | The exercises were useful towards learning the material because ***we got to look at the problems to figure out ourselves*** rather than just following along in class. [Fall '16, health science major] |
| | ***Yes, I actually did.*** [Spring '17, health science major; answer to Question #2 was "No sir"] |

**FIG. 3**. (Top) The portion of the end-of-semester feedback survey that pertains to the usefulness of the pre-laboratory group problem solving exercise. (Bottom, left) The three learning orientation categories, including criteria that served to classify student responses as belonging to a respective category. (Bottom, right) Sample self-reported statements of students in response to Question #1, including highlighted language that meet the criteria of the orientation category in which the student is classified. Basic information about each student, and answers to Question #2 where pertinent, are included in brackets.

At the end of the exercise, the instructor reviewed the problem solution to ensure that everyone understood how to think about the problem while solving it. Students also had to complete, and submit for completion credit, a metacognitive self-monitoring reflection on individual strengths and struggles of attempting the problem. Fig. 2 shows the scoring rubric that each student was required to fill out with respect to their experience in attempting to solve the problem with lab partners. The rubric is a slight adaptation from scaffolding provided in a previous experiment on self-monitoring, specifically self-diagnosis of quiz errors, conducted by Yerushalmi et al.[40,41]

### C. Data Collection

In addition to course grades, the researchers recorded pre-post survey data from the FCI[43] and CLASS[44] for analysis, with the pretests occurring during the first laboratory session of the semester and the posttests occurring during the final laboratory session of the semester. For the pretest, the instructor asked students to consider their views of learning the most recent introductory physical science course they had taken instead (e.g. high school chemistry or middle-school physical science). The majority of students in any given classroom typically had never taken a high school physics course.

Also collected was a post-test survey which asked students about the usefulness of the problem-solving exercise. Figure 3 displays a copy of the post-test survey questions that are pertinent to the study, as well as a few sample student responses from three of the selected semesters, organized into each of three learning orientations as determined by the researchers using inter-rater reliability. The central question of the survey was as follows: "In what ways did you find the problem-solving exercise useful?" Answers to this question were interpreted in terms of Framework-orientation and/or Performance-orientation. Results were confirmed using inter-rater reliability check by two researchers, who independently categorized student responses without any reference to student IDs before comparing results. In a few specific cases where a student response could feasibly be interpreted as having more than one orientation, the researchers would then consider the student's response to a secondary question in terms of learning orientations: "Do you have any suggestions to make this exercise more useful toward learning the material in the course?" Consideration of this secondary question typically resolved uncertainties regarding the student's primary learning orientation. If the student's learning orientation was not made clear by either the primary or secondary question's response, then the student would be considered to not have an achievement goal in mind (i.e. "Vague" orientation). Initial agreement per semester ranged between 80% and 90%, and the researchers were able to quickly resolve differences in categorization in the remaining cases.

76 students' responses out of 218 actually responded to the survey question in terms of attempting to master all or part of a problem-solving framework (hence the term "Framework-oriented," i.e. mastery-oriented within the scope of mastering a problem-solving framework according to the described in-class exercise). In contrast, 79 students' responses were primarily oriented with regard to how the exercise would help them perform better in other aspects of the course: exams, homework, or the lab experimental activity that followed the exercise (hence the term "Performance-oriented"). The remaining 63 students did not directly answer the question (hence the term "Vague-oriented"); their responses were either indirectly related to the problem-solving exercise (e.g. they liked working in groups and getting feedback from their lab partners; it was preferable to another aspect of the course), or unrelated entirely (e.g. no response, comments about other aspects of the course).III. Results

Here in the Results section, we will discuss data analysis in the following order. First, we briefly examine the distribution of learning orientations among each set of majors. Second, we analyze the pre-post CLASS problem solving items, both with respect to the learning orientations (in order to specifically address research question 1), and with respect to choice of major (as part of the response to research questions 2 and 3). Third, we analyze the pre-post overall FCI and CLASS results, as well as overall course grade results, both by learning orientation and by choice of major, also as a response to research questions 2 and 3.

## IV. RESULTS

### A. Distribution of learning orientations among each set of majors

Table I presents the 218 students' distribution by learning orientation ("Framework," "Performance," and "Vague"), as determined by analysis of survey responses as discussed in Figure 3, and by choice of major. Biology majors and health science majors appear to have very similar distributions according to learning orientation. The sample of other natural science majors is relatively small, so references to this sample are of secondary importance to the study; however, note that almost all of the other natural science majors were either framework-oriented or vague-oriented.

**TABLE I.** Students categorized into problem solving exercise orientations, per end-of-semester survey responses.

| # Students | Biology | Other Natural Science | Health Science | Non-Science | Total # by Orientation |
|---|---|---|---|---|---|
| Framework | 36 | 5 | 32 | 3 | 76 |
| Performance | 30 | 18 | 28 | 3 | 79 |
| Vague | 25 | 12 | 25 | 1 | 63 |
| **Total # by Major** | 91 | 35 | 85 | 7 | 218 |

### B. Comparison of learning orientation categories to CLASS Problem-Solving cluster pre and post-test results

To address the validity of the learning orientation categorizations more specifically, we first focus on the problem solving (PS) item clusters within the CLASS survey. Table II features pretest scores and normalized gains on the CLASS item clusters that involve problem solving ("General," i.e. PS-G; "Confidence," i.e. PS-C; and "Sophistication," i.e. PS-S). Item clusters that regard conceptual understanding (overall, i.e. CU, and "Applied," i.e. ACU) are also shown in Table II. Each individual student's pretest and posttest scores were calculated in terms of percentage of expert-like responses (i.e. either "Agree" or "Strongly Agree" for an expert-like statement, or conversely "Disagree" or "Strongly Disagree" for a novice-like statement); CLASS questions for which there was not an expert-like opinion are omitted. For Table II, and all ensuing tables in which one-way ANOVA comparisons between groups are reported, Levene's test[45] was conducted to check for equality of variances between groups on each comparison, and one-way ANOVA comparisons were conducted accordingly. In the vast majority of cases, two compared groups had approximately equal variances ($p > 0.05$); equality of variance or lack thereof did not influence any given comparison's p-values.

On the pretest, health science majors were more novice-like than were biology majors on all clusters except for the PS-Confidence cluster ($p > 0.50$, PS-C; $p < 0.05$, all other comparisons), and more novice-like than were other natural science majors on most pretest comparisons ($p > 0.10$, PS-C; $p = 0.07$, PS-S; $p < 0.001$, all other comparisons). The gains are all statistically insignificant as well, both in terms of remaining within standard error of zero gain and in terms of relative differences; this indicates that health science majors remain more novice-like than do natural science majors. By learning orientation, the overall pretest pattern holds for all problem solving and conceptual understanding item clusters, with no statistically significant differences between any groups in any item cluster reported in Table II. However, the framework-oriented students have statistically higher gains across all item clusters than do the vague-oriented students ($p < 0.02$, all comparisons); with the exception of the PS-C cluster, framework-oriented students also have stronger gains than do performance-oriented students ($p > 0.3$, PS-C; $p < 0.01$, all other comparisons). Performance-oriented and vague-oriented students were either borderline-significant ($p < 0.1$, PS-C and ACU) or statistically similar ($p > 0.1$, all other comparisons).

Overall, it appears that averaged individual gains on the CLASS problem solving item clusters match the expectations for the learning orientation categories, namely that Framework-oriented students on average exhibit a more expert-like normalized gain on the CLASS than do the Performance-oriented or Vague-oriented students. Indeed, additional categories of the

**TABLE II.** Average pretest scores and gains for students by major on CLASS item clusters, in terms of percentage of answers that are expert-like. Standard errors are presented in the brackets (e.g. "(3)" = +/- 3%). Bold numbers indicate statistical significance within a row, either by orientation or by choice of major; italicized numbers similarly indicate borderline statistical significance. Please see text discussion on statistical differences for further detail.

| Group | | All | By orientation | | | By choice of major | | |
|---|---|---|---|---|---|---|---|---|
| | | | Framework | Performance | Vague | Biology | Other NS | Health Sci |
| Cluster | n | 218 | 76 | 79 | 63 | 91 | 35 | 85 |
| Overall | Pre (%) | 57 (1) | 57 (2) | 57 (2) | 56 (2) | 60 (1) | 61 (3) | **51** (2) |
| | Gain (<g>) | +.02 (.02) | **+.11** (.03) | -.03 (.03) | -.03 (.03) | +.03 (.03) | +.04 (.05) | -.00 (.03) |
| **PS-G** | Pre (%) | 63 (2) | 63 (3) | 63 (3) | 62 (3) | 65 (2) | 67 (4) | **58** (2) |
| | Gain (<g>) | -.01 (.03) | **+.14** (.05) | **-.02** (.05) | **-.08** (.05) | +.01 (.05) | +.12 (.08) | -.02 (.05) |
| **PS-C** | Pre (%) | 64 (2) | 64 (5) | 64 (4) | 65 (4) | 64 (3) | 69 (4) | 61 (3) |
| | Gain (<g>) | -.01 (.04) | *+.11* (.07) | -.02 (.07) | *-.13* (.06) | +.05 (.06) | +.10 (.09) | -.07 (.06) |
| **PS-S** | Pre (%) | 43 (2) | *40* (3) | 42 (4) | *47* (4) | *45* (3) | 56 (5) | **35** (2) |
| | Gain (<g>) | -.10 (.03) | **+.05** (.05) | -.14 (.06) | -.24 (.05) | -.05 (.05) | -.17 (.09) | -.11 (.05) |
| CU | Pre (%) | 54 (2) | 53 (4) | 53 (3) | 57 (4) | 57 (2) | 66 (5) | **46** (3) |
| | Gain (<g>) | +.02 (.03) | **+.18** (.05) | -.05 (.06) | -.08 (.05) | +.03 (.05) | -.01 (.09) | +.01 (.05) |
| ACU | Pre (%) | 42 (2) | 41 (3) | 40 (3) | 45 (3) | *44* (2) | 52 (4) | **36** (2) |
| | Gain (<g>) | -0.06 (.03) | **+.08** (.05) | -.08 (.05) | -.22 (.06) | +.01 (.04) | -.13 (.08) | *-.11* (.05) |

**TABLE III.** Average FCI pretest scores, posttest scores, and individual normalized gains for students, by major and by orientation, in terms of percentage of correct answers. See Table I for sample sizes. Standard errors are presented in parentheses next to reported values (e.g. "(3)" = +/- 3%).

| **Pretest scores (%)** | Biology | Other Natural Science | Health Science | Total by Orientation |
|---|---|---|---|---|
| Framework | 24 (2) | 35 (8) | 21 (2) | 24 (1) |
| Performance | 25 (2) | 29 (4) | 23 (2) | 25 (1) |
| Vague | 30 (3) | 40 (7) | 24 (2) | **29** (2) |
| Total by Major | **26** (1) | **34** (2) | **22** (3) | 26 (1) |

**Statistical Significances for Pretest:** Across Total by Major row, all three types of majors; across Total by Orientation column, Vague pretest is higher than Framework pretest and borderline higher than Performance pretest.

| **Posttest scores (%)** | Biology | Other Natural Science | Health Science | Total by Orientation |
|---|---|---|---|---|
| Framework | 39 (3) | 45 (9) | 32 (2) | 37 (2) |
| Performance | 37 (3) | 46 (5) | 31 (2) | 36 (2) |
| Vague | 40 (3) | 50 (6) | 32 (2) | 39 (2) |
| Total by Major | **39** (3) | **47** (4) | **31** (2) | 37 (1) |

**Statistical Significances for Posttest:** Across Total by Major row, all three types of majors.

| **Avg. normalized gains** | Biology | Other Natural Science | Health Science | Total by Orientation |
|---|---|---|---|---|
| Framework | +.21 (.03) | +.17 (.09) | +.15 (.03) | **+.18** (.02) |
| Performance | +.16 (.03) | +.25 (.04) | +.08 (.03) | +.14 (.02) |
| Vague | +.13 (.04) | +.15 (.04) | +.11 (.02) | **+.13** (.02) |
| Total by Major | +.17 (.03) | +.20 (.03) | **+.11** (.02) | +.17 (.01) |

**Statistical Significances for Individual Normalized Gains:** Across Total by Major row, Health Science gains are lower than other groups' gains; across Total by Orientation column, Framework gains are higher than Vague gains.

CLASS, as well as the overall result on all CLASS items with an expert-novice difference in possible answer responses, appear to extend a more expert-like set of gains for Framework-oriented students.

### C. Overall FCI pre and post-test results

Table III shows results on the FCI for the student population, both by choice of major and by learning orientation. Table III shows a statistically significant difference between all three sets of majors on the pretest ($p < .05$, each comparison); the other Natural Science majors had the best overall FCI pretest score, followed by Biology majors, and Health Science majors had the lowest FCI pretest scores. The differences between majors remains on the posttest scores ($p < .03$, all comparisons); in fact, the normalized gain for health science majors is statistically lower than respective gains for both groups of natural science students ($p < .02$, each comparison). It therefore appears that on average, health science majors begin the course with a disadvantage in terms of force and motion concepts, respective to natural science majors; the disadvantage appears to persist through the end of the course as well, according to the posttest results.

In terms of choice of learning orientation, the vague-oriented students start instruction with a slight advantage over framework-oriented students ($p < .03$) and a borderline advantage over performance-oriented students ($p = 0.11$). However, this advantage does not exist on the posttest scores ($p > 0.30$, all comparisons), and framework-oriented students' gains are statistically larger than vague-oriented students' gains ($p < .05$). In other words, learning orientation does not appear to show any differences overall on the FCI posttest.

While the gains appear relatively poor in comparison to typical normalized gain reports from introductory physics classrooms,[46] the effect size[47,48] from pretest to posttest over all 218 students appears strong (Cohen's d = +0.78), and ranges from d = 0.6 to d = 0.9 when looking specifically at each respective choice of major and each respective learning orientation. This result is likely due to the pretest scores on the FCI, which are close to the random-chance level of 20% (i.e. the odds of randomly choosing the correct answer out of 5 choices for each question).

### D. Overall CLASS pre and post-test results

Similar to Table III, Table IV shows pretest and posttest scores, as well as average individual normalized gains, on the overall CLASS survey. In terms of choice of major, the health science majors begin the course in a more novice-like state than do biology majors ($p <$

0.0001) and other natural science majors ($p < 0.002$). There is no significant difference between any of the three sets of majors in terms of gains ($p > 0.40$, all comparisons), and the gains themselves are all within standard error of zero net gain. Therefore, the health science majors remain more novice-like than both sets of natural science majors ($p < 0.01$, both comparisons) at the end of the course.

In terms of learning orientation, there is no difference between learning orientations in attitudes towards physics on the pretest ($p > 0.70$, all comparisons). On the posttests, however, the framework-oriented students are significantly more expert-like than are the other two learning orientations ($p < 0.02$, both comparisons); this is also true for gains ($p < 0.01$, both comparisons), where the framework-oriented students trend in a more expert-like direction, and the other students exhibit a slight overall decline. Also of note is that the framework-oriented students' advantage in effect size[47,48] on the FCI from pre to post is relatively small ($d_F - d_P = 0.13$, $d_F - d_V = 0.31$).

Looking more closely within the subgroups of learning orientations by choice of majors, the trend of Framework-oriented students experiencing more expert-like shifts appears strongest within health science majors, to the point of statistical significance ($p < 0.05$, both respective comparisons with Performance-oriented and Vague-oriented health science majors). The trend can be seen within biology majors as well, but it is not statistically significant ($p > 0.20$, both respective comparisons).

Learning orientations were specifically defined with respect to students' views specifically regarding the pre-laboratory problem-solving exercise, not towards the course as a whole. Yet, as Table IV indicates, the overall CLASS survey seems to indicate the framework-oriented students gain a more expert-like view of the overall course, not just of the problem-solving exercise. Normalized gains on the CLASS in the literature typically expect a slightly more novice-like, i.e. unfavorable, shift from pre to post.[49,50] However, recent research-based interventions can indeed produce a more

**TABLE IV.** Average overall CLASS pretest scores, posttest scores, and gains for students, by major and by orientation, in terms of percentage of answers that are expert-like. See Table I for sample sizes. Standard errors are presented in parentheses next to reported values (e.g. "(3)" = +/- 3%).

| **Pretest scores (%)** | Biology | Other Natural Science | Health Science | Total by Orientation |
|---|---|---|---|---|
| Framework | 59 (3) | 57 (6) | 54 (2) | 57 (2) |
| Performance | 58 (3) | 65 (4) | 50 (3) | 57 (2) |
| Vague | 63 (2) | 56 (6) | 48 (3) | 56 (2) |
| Total by Major | **60** (1) | **61** (2) | **51** (3) | 57 (1) |

**Statistical Significances for Pretests:** Across Total by Major row, Health Science is lower than Biology and Other Natural Science.

| **Posttest scores (%)** | Biology | Other Natural Science | Health Science | Total by Orientation |
|---|---|---|---|---|
| Framework | 62 (3) | 59 (8) | 66 (3) | **61 (2)** |
| Performance | 57 (3) | 61 (5) | 46 (3) | **54 (2)** |
| Vague | 59 (3) | 59 (5) | 44 (3) | **53 (2)** |
| Total by Major | **59** (2) | **61** (3) | **50** (2) | 56 (1) |

**Statistical Significances for Posttests:** Across Total by Major row, Health Science is lower than Biology and Other Natural Science; across Total by Orientation column, Framework is higher than Performance and Vague.

| **Avg. normalized gains** | Biology | Other Natural Science | Health Science | Total by Orientation |
|---|---|---|---|---|
| Framework | +.08 (.05) | +.19 (.16) | **+.11 (.05)** | **+.11 (.03)** |
| Performance | +.01 (.06) | -.02 (.05) | **-.06 (.08)** | **-.03 (.03)** |
| Vague | -.02 (.06) | +.08 (.07) | **-.08 (.04)** | **-.03 (.03)** |
| Total by Major | +.03 (.03) | +.04 (.05) | -.00 (.03) | +.02 (.02) |

**Statistical Significances for Normalized Gains:** Across Total by Orientation column, Framework gains are higher than Performance and Vague gains; across Health Science column, Framework gains are higher than Performance and Vague gains.

expert-like, favorable shift.[28,51] The student sample in Table IV was overall within statistical error of zero gain, albeit with a slightly positive favorable shift; however, the learning orientation categories offer a means of differentiating between a group of students who on average become more expert-like overall in attitudes in introductory physics, and a group of students who do not on average become more expert-like overall.

### E. Overall course grade results

Interpreting the results so far, health science majors appear to start the course at a lower level of conceptual understanding, and with a more novice-like view of learning physics and physical science, than do natural science majors. Moreover, they do not catch up with biology majors or other natural science majors during the course. There are similar concerns by learning orientation, as framework-oriented students appear to be the only group that has positive gains on the CLASS survey across the item clusters in Table II, while performance-oriented students and vague-oriented students slightly trend in the more novice-like direction. As a result, a concern arises that health science majors, and alternately non-framework-oriented students, may also perform more poorly in the overall course.

Table V shows the overall course GPAs by choice of major and by learning orientation. In both cases, each student's course GPA is interpreted as follows: A = 4.0, B = 3.0, C = 2.0, and D = 1.0. When considering students by choice of major, Health Science majors average almost two thirds of a letter grade below Biology majors. In addition, the Biology majors appear to have performed significantly better than did other, non-lif-science Natural Science majors.

In contrast, when considering students by learning orientation groups in Table V, there is only a borderline difference between vague-oriented students and the other orientations. There does not appear to be a correlation between learning orientation and choice of major in this regard, either. A check of learning orientations by choice of major showed no difference within each major by learning orientation ($p > 0.05$, all comparisons within respective major groups), but a consistent difference in favor of biology majors over health science majors for each learning orientation ($p < 0.05$, all comparisons within respective orientation groups).

**TABLE V.** Average course grade for major groups, with standard error for each group and p-values between groups.

| Student Group | Average GPA | SE | | | |
|---|---|---|---|---|---|
| All Students (218) | 2.87 | 0.06 | | | |
| Biology (91) | 3.18 | 0.08 | | Biology vs. Health | **<0.00001** |
| Health (85) | 2.58 | 0.10 | **p-values by Major** | Biology vs. Other NS | **<0.02** |
| Other NS (35) | 2.80 | 0.14 | | Health vs. Other NS | 0.22 |
| Framework (76) | 2.91 | 0.10 | | Framework vs. Performance | 0.63 |
| Performance (79) | 2.97 | 0.09 | **p-values by Orientation** | Framework vs. Vague | 0.15 |
| Vague (63) | 2.68 | 0.12 | | Performance vs. Vague | *0.05* |

### V. LIMITATIONS OF STUDY

We must note that the study contains limitations due to the unexpected source of learning orientations as self-expressed by the students, namely the feedback survey. This survey was originally designed simply to get student feedback about the problem solving exercise, for the purposes of adjusting the exercise as needed; it was not designed a priori to be a validated data collection instrument, and therefore required an unorthodox approach towards checking validity. While the authors' research question was satisfied, i.e. validity in terms of correlation with expected CLASS results was satisfied, the validity in terms of defining the learning orientation categories themselves remains somewhat weak. Students chose to answer the principal survey question in terms of mastery and performance achievement goals, similar to the two primary learning orientations detected by Hazari et al. Therefore, the degree to which students did express achievement goals (and therefore the degree to which the authors could use students' responses to define learning orientation categories) is limited to single-item short-answer statements, as opposed to the more meticulous study design for a larger, longitudinal study by Hazari et al. Indeed, not all students expressed either Performance or Framework orientations; the Vague group in particular displayed a mixture of process-oriented responses (most prominently, that some students liked working in groups) and impertinent or null responses.

Therefore, in order to further explore and check the robustness to this paper's results in future studies, a more robust feedback survey must be generated with deliberate hypotheses in mind. For example, in the Hazari et al. study, the authors explicitly designed a learning-orientations survey from hypotheses generated from in-depth interviews, using multiple statistical tests to confirm the validity of the survey's results. A similar approach may be taken within the scope of introductory physics courses for IPLS-esque course populations. This will serve to obtain students' learning orientations with more thorough expression than can be had by a single-items short answer response. Additional dimensions of achievement goal theory that have not been considered within the scope of this study, e.g. approach vs. avoidance motivations,[3] may increase the depth and rigor of a future study on students' learning orientations.

## VI. CONCLUSIONS AND FUTURE DISCUSSION

### A. Validity of the Learning Orientation Categories between Students: Pre-Post Attitudinal Shifts on Problem Solving

Analysis of pre-post attitudinal shifts on CLASS problem-solving items appears to show a robust correlation with empirically determined learning orientation categories. The Framework-oriented students (who valued the problem-solving exercise in terms of a learning a problem-solving framework for its own sake, and more generally on improving their current state of knowledge) have a statistically significant expert-like shift, while Performance-oriented students and Vague-oriented students experience a slight novice-like shift. It can therefore be inferred that mastery-oriented (i.e. Framework-oriented) students, who expressed a desire to improve their current state of learning and understanding physics problem solving, could on average be expected to become more expert-like on attitudes towards physics problem solving than other students who did not express a desire to improve their understanding of physics problem solving.

Usage and future refinement of student learning orientations within an introductory physics course, in terms of reflecting a robust difference on CLASS problem solving clusters and on the CLASS as a whole, appears to offer a closer examination of a trend within PER regarding pre-post CLASS attitudinal shifts. A slight novice-like shift, as seen by the Performance-oriented and Vague-oriented students, has been reported as common for traditional-format course populations in PER literature.[49,50] An overall expert-like shift in attitudes, as experienced by the Framework-oriented students, is not uncommon for overall results in courses that have undergone research-based reforms.[28,51] In this way, a difference in learning orientations (and therefore achievement goals) between subgroups of a student population appears to correlate to a clear difference in attitudinal shifts between said subgroups. The result can thus be interpreted as showing that students' achievement goals (or lack thereof) potentially have a direct effect on whether or not a research-based exercise, performed throughout an introductory physics course, will correlate with an expert-like shift in students' attitudes towards physics.

### B. Comparison between Learning Orientations on General Measurements for the Entire Course

The results between learning orientations regarding the CLASS problem solving clusters appears to extend to overall pre-post CLASS results. Framework-oriented students, who expressed a genuine interest in developing and mastering problem-solving skills, experienced an overall expert-like shift, while Performance-oriented students, who were more interested in how problem solving would help them perform better in the overall course, experienced an overall novice-like shift. What is striking is that the two orientation groups began the course with nearly identical CLASS pretest scores on average; this may reflect the fact that the vast majority of students in the course have never taken physics prior to this introductory physics course, and so students needed to undergo instruction to differentiate themselves on the material. However, this difference in overall CLASS pre-post results does not appear to extend to FCI pre-post results or overall average course performance. Framework-oriented and Performance-oriented students have virtually identical course performances on average, while Vague-oriented students are marginally worse than the other two groups.

The lack of correlation of learning orientations (or for that matter, achievement goals) to overall course grades and FCI pre-post results suggests at least two possible interpretations. First, with regard to course performance, a single course in introductory, algebra-based physics is not necessarily a large enough scope in order to see an effect of learning orientations on overall course performance or conceptual understanding. The lack of effect on course performance in this study is in contrast with the Hazari et al. study of a longitudinal study of graduate-level physical science students, in which they did find a difference in career success rates.[13] Performance-oriented students might indeed be expected to perform as well as Framework-oriented students perform, on average, within the scope of a single course. Therefore, similar analysis for a longitudinal study on a two- or three-course introductory physics sequence may be necessary to see if mastery-oriented students have better rates of success as more content is covered. However, this approach has its limits, as non-physical-science major tracks typically do not need many physics courses outside the introductory sequence at many institutions. To account for such limitations, it may be necessary to track students' achievement goals through their respective major tracks, and compare findings to similar findings from their introductory physics courses.

Second, with regard to FCI pre-post results, the issue may have to do with a lack of practice in conceptual thinking about physical science. The vast majority of students in the student sample reported that they had never taken a regular physics class in high school or at the college level prior to beginning the course. In other words, the exposure of most students to physics prior to beginning the course was likely limited to middle-school-level physical science. This is in contrast to many studies of calculus-based introductory physics courses, in which many students typically have taken algebra-based or even calculus-based physics in high school prior to enrolling in the course. The average FCI pretest score over the entire student sample was relatively close to a random-guess score of 20% (i.e. one has a 20% chance of picking the correct answer when randomly choosing from 5 choices). The authors of the FCI note that students' common sense misconceptions must be accounted for and confronted by an introductory physics course in order for the course to be effective,[43,54] and that the FCI is intended as a measurement of belief systems (between Newtonian concepts and common-sense misconceptions).[43] Here, a simple lack of preparation of students in the algebra-based introductory physics course may account for a very poor FCI pretest showing and pre-post gains that appear unrelated to learning orientations.

The Vague-oriented students, like the Performance-oriented students, appeared to have slightly novice-like attitudinal shifts. However, for certain item clusters on the CLASS, the Vague-oriented students are noticeably more novice-like than are Performance-oriented students. In addition, the Vague-oriented students performed marginally worse on average in the overall course than did students whose orientations reflected an achievement goal (Framework-oriented or Performance-oriented). This presents a concern in that roughly 30% of the sampled students were Vague-oriented.

One potential consideration may simply have to do with lack of interest in the course material, particularly among the subset of Vague-oriented students who did not express either an achievement goal or even a focus on the group process. A simple lack of interest in physics from students has been identified as a persistent problem among secondary school students,[52,53] in a way that frustrates pedagogical development and goals; this issue may extend to non-physics science majors in introductory algebra-based physics courses. This would not be true for all Vague-oriented students, though, as several individual students in this group performed very well overall in the course.

### C. Degree of Influence of Learning Orientations in Comparisons between Biology and Health Science Majors

Biology majors appear to have a strong advantage over health science majors in the course. They experience stronger conceptual gains on the FCI, a relatively more expert-like attitude towards physics on the CLASS, and a much stronger average overall course performance. Comparisons of learning orientation subgroups within each major appear to have minimal influence on the overall difference between the two sets of majors; the relative distribution of learning orientation populations within each set of majors is virtually identical. Patterns regarding relative course performance between learning orientations within each major (i.e. within biology and also within health science) also appear to be somewhat similar for each set of majors. The only noteworthy difference appears to be that health science majors who are not Framework-oriented appear to perform significantly worse than their Framework-oriented counterparts; this is not true for biology majors. However, this relative benefit for Framework-oriented health science majors, with respect to other health science majors, does not translate to an overall benefit when compared to biology majors.

### D. Future Directions of Study

It is clear that there is a difference in attitudinal shifts, as well as conceptual gains and course performance, between biology majors and health science majors, and that furthermore this difference between sets of majors occurs regardless of learning orientations (insofar as learning orientations could be established with responses to a limited feedback survey). Admittedly, this may not be a concern for IPLS courses with a monolithic life science population (e.g. courses in which pre-medical biology majors are typically close to 100% of enrollment). However, it does present a concern for institutions similar to the study's host university, in which approximately equal amounts of biology majors and health science majors take the same introductory physics course and clearly experience different average results within the course. A tempting solution is to offer two different sets of IPLS-oriented introductory physics courses, and this solution may be straightforward for larger physics departments who can spare the additional faculty contact hours for both sets of courses. However, not all institutions who experience this mixture of life science majors have sufficient resources to make this change.

As mentioned within the Limitations section, future explorations for this study may require a more methodically constructed feedback survey based upon well-established research hypotheses, as with the Hazari et al. study, but within the scope of an introductory physics course. Other considerations can be made in this regard pertaining to other well-defined classifications within mastery and performance achievement goals. For example, approach vs. avoidance motivations have long been recognized within psychological applications,[3] and both have been recognized to exist respectively within mastery goals and within performance goals. In other words, students may express mastery-related approach motivations and mastery-related avoidance motivations, or they may express similar performance-related motivations. A differentiation between achievement goal focus, process focus, and lack of focus may also prove fruitful in advancing the understanding of student motivations.

Another possible future study entails the consideration of a curriculum that explicitly endorses mastery-oriented goals. A recent experiment[55] sought to explicitly alter achievement goals with a mastery-structured learning environment; the results suggest further considerations in studying life science majors' achievement goals within introductory physics. For example, the researchers of this study identified two "contingencies of self-worth" as being co-variant with performance goal orientations.

If achievement goals, by way of learning orientations, do not explain the difference between biology majors and health science majors, then other considerations may instead explain this difference. One potential explanation is in terms of disciplinary norms within the two sets of science majors,[41] specifically how differences in discipline may more deeply describe differences in course outcomes between health science students and natural science students. Another potential explanation is the analysis of pre-professional status (e.g. students who declare a pre-medical or pre-physical therapy track) as a potential variable. The researchers did examine the pre-professional status of the current student sample as a potential variable; results are currently inconclusive and should be examined further in future studies.

Finally, it may be possible to identify a subgroup of Vague-oriented students regarding lack of interest, as differentiated from other Vague-oriented students. If so, then a variable to account for this may be identified in future analyses of student populations. To define this variable properly, it may be necessary to consider current and future studies into high-school populations[52,53] to address the lack of interest prior to enrolling in a university-level introductory physics course.

## VI. ACKNOWLEDGEMENTS

The authors would like to thank the Editor, Editorial Board, and Referees of this manuscript for their thorough and helpful advice in the revision process. We also thank the UCA Society of Physics Students for their respective contributions in collecting data and discussions about data analysis. We also thank E. Redish and B. Geller for useful discussions about this project and for assistance with existing IPLS literature, and T. Nokes-Malach for assistance with literature in cognitive psychology that pertains to achievement goal theory. The University of Central Arkansas Sponsored Programs Office and Department of Physics and Astronomy provided funding for this project.